\documentclass{SciPost}

\hypersetup{
    colorlinks,
    linkcolor={red!50!black},
    citecolor={blue!50!black},
    urlcolor={blue!80!black}
}

\usepackage[bitstream-charter]{mathdesign}
\DeclareSymbolFont{usualmathcal}{OMS}{cmsy}{m}{n}
\DeclareSymbolFontAlphabet{\mathcal}{usualmathcal}

\fancypagestyle{SPstyle}{
\fancyhf{}
\lhead{\colorbox{scipostblue}{\bf \color{white} ~SciPost Physics Proceedings }}
\rhead{{\bf \color{scipostdeepblue} ~Submission }}

\fancyfoot[C]{\textbf{\thepage}}
}

\usepackage{booktabs} 
\usepackage{multirow} 
\usepackage{xspace} 
\def\LOone{${\rm LO}_1$\xspace}
\def\LOtwo{${\rm LO}_2$\xspace}
\def\LOthree{${\rm LO}_3$\xspace}
\def\LOfull{${\rm LO}$\xspace}
\def\NLOone{${\rm NLO}_1$\xspace}
\def\NLOtwo{${\rm NLO}_2$\xspace}
\def\NLOthree{${\rm NLO}_3$\xspace}
\def\NLOfour{${\rm NLO}_4$\xspace}
\def\NLOqcd{${\rm NLO}_{\rm QCD}$\xspace}
\def\NLOfull{${\rm NLO}$\xspace}
\def\NLOprd{${\rm NLO}_{\rm prd}$\xspace}

\def\Recola{\textsc{Recola}\xspace}

\begin{document}

\pagestyle{SPstyle}

\begin{center}{\Large \textbf{\color{scipostdeepblue}{
Complete NLO corrections to $t\bar{t}\gamma$ and $t\bar{t}\gamma\gamma$\\
}}}\end{center}

\begin{center}\textbf{
Daniel Stremmer\textsuperscript{1,2$\star$},
}\end{center}

\begin{center}
{\bf 1} Institute for Theoretical Particle Physics and Cosmology, RWTH Aachen University, \\D-52056 Aachen, Germany
\\
{\bf 2} Institute for Theoretical Particle Physics,
Karlsruhe Institute of Technology, \\D-76128 Karlsruhe, Germany
\\[\baselineskip]
$\star$ \href{mailto:email1}{\small daniel.stremmer@kit.edu}
\end{center}

\definecolor{palegray}{gray}{0.95}
\begin{center}
\colorbox{palegray}{
  \begin{tabular}{rr}
  \begin{minipage}{0.36\textwidth}
    \includegraphics[width=60mm,height=1.5cm]{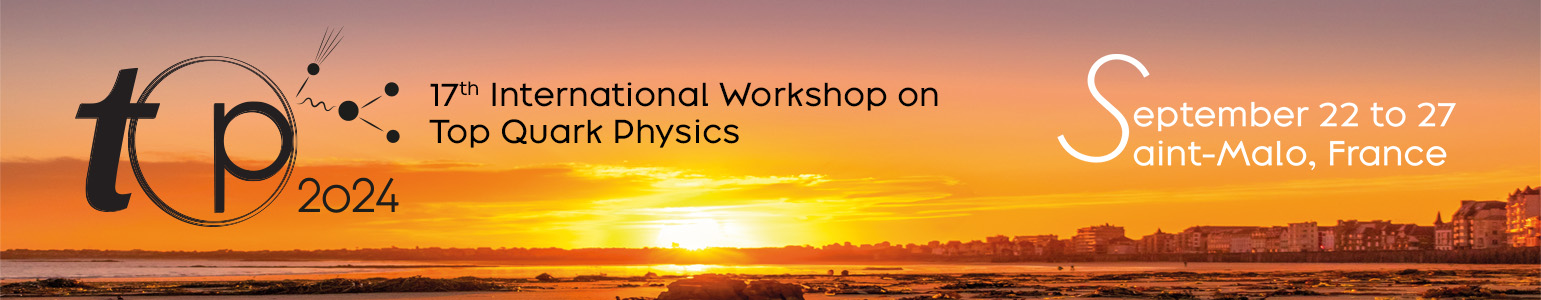}
  \end{minipage}
  &
  \begin{minipage}{0.55\textwidth}
    \begin{center} \hspace{5pt}
    {\it The 17th International Workshop on\\ Top Quark Physics (TOP2024)} \\
    {\it Saint-Malo, France, 22-27 September 2024
    }
    \doi{10.21468/SciPostPhysProc.?}\\
    \end{center}
  \end{minipage}
\end{tabular}
}
\end{center}

\section*{\color{scipostdeepblue}{Abstract}}
\textbf{\boldmath{%
In this contribution we discuss recent progress in associated top-quark pair production with one or two isolated photons, $pp\to t\bar{t}\gamma(\gamma)$. The focus is the simultaneous inclusion of higher-order effects and photon radiation in the production of the top-quark pair and in the decay processes. This allows us to quantify the importance of photon radiation in decay processes and the size of the so-called complete NLO corrections in realistic final states.
}}
\begin{flushright} P3H-24-095, TTK-24-55\end{flushright}

\vspace{\baselineskip}

\noindent\textcolor{white!90!black}{%
\fbox{\parbox{0.975\linewidth}{%
\textcolor{white!40!black}{\begin{tabular}{lr}%
  \begin{minipage}{0.6\textwidth}%
    {\small Copyright attribution to authors. \newline
    This work is a submission to SciPost Phys. Proc. \newline
    License information to appear upon publication. \newline
    Publication information to appear upon publication.}
  \end{minipage} & \begin{minipage}{0.4\textwidth}
    {\small Received Date \newline Accepted Date \newline Published Date}%
  \end{minipage}
\end{tabular}}
}}
}




\section{Introduction}
\label{sec:intro}

Among the associated production processes of a top-quark pair with a vector boson, $pp\to t\bar{t}+V$ (with $V=\gamma,W^{\pm},Z$), the $pp\to t\bar{t}\gamma$ process is the dominant one at the LHC and was first observed by the ATLAS collaboration at the center-of-mass energy of $\sqrt{s}=7 ~{\rm TeV}$ \cite{ATLAS:2015jos}. This production process inherits many unique challenges with respect to other associated $t\bar{t}$ processes. In particular, a large fraction of photon radiation originates from top-quark decays, which makes the modelling of this process more difficult. In addition, the use of realistic photon isolation conditions leads to further difficulties in theoretical predictions. Finally, the $pp\to t\bar{t}\gamma$ process can be used to probe the $t-\gamma$ coupling.

On the other hand, the $pp\to t\bar{t}\gamma\gamma$ process was not observed yet and is even more affected by these problems due to the additional photon. In addition, this process is part of the irreducible background to the $pp\to t\bar{t}H$ process in the $H\to\gamma\gamma$ decay channel, which was the first single-channel observation of $t\bar{t}H$ \cite{ATLAS:2020ior,CMS:2020cga}.

In this report we discuss improved theoretical predictions for the two processes $pp\to t\bar{t}\gamma$ and $pp\to t\bar{t}\gamma\gamma$ where we consistently include photon radiation as well as all LO and NLO contributions in the production and the decays of the two top quarks. In addition, we discuss the prompt photon distribution for the $pp\to t\bar{t}\gamma\gamma$ process by dividing the full calculation into different resonant contributions based on the origin of the radiated photons in the decay chain. All results shown in this report are taken from Refs. \cite{Stremmer:2023kcd,Stremmer:2024ecl}, where further information can be found.

\section{Process definition and computational setup}
In order to study the significance of photon radiation from the decay processes, we include the decays of top quarks and $W$ bosons employing the narrow-width approximation and consider the di-lepton top-quark decay channel. In this case we can split the full calculation of $pp\to t\bar{t}\gamma\gamma$ into three resonant contributions based on the origin of the photons according to
\begin{equation}
\label{eq_res}
\begin{split}
    d\sigma_{\rm Full}&=\overbrace{d\sigma_{t\bar{t}\gamma\gamma}\times\frac{d\Gamma_t}{\Gamma_t}\times\frac{d\Gamma_{\bar{t}}}{\Gamma_t}}^{\sigma_{\rm Prod.}} + \overbrace{d\sigma_{t\bar{t}\gamma}\times\left( \frac{d\Gamma_{t\gamma}}{\Gamma_t}\times\frac{d\Gamma_{\bar{t}}}{\Gamma_t} + \frac{d\Gamma_t}{\Gamma_t}\times\frac{d\Gamma_{\bar{t}\gamma}}{\Gamma_t} \right)}^{\sigma_{\rm Mixed}}
    \\&
    +\underbrace{d\sigma_{t\bar{t}}\times\left( \frac{d\Gamma_{t\gamma\gamma}}{\Gamma_t}\times\frac{d\Gamma_{\bar{t}}}{\Gamma_t} + \frac{d\Gamma_t}{\Gamma_t}\times\frac{d\Gamma_{\bar{t}\gamma\gamma}}{\Gamma_t} + \frac{d\Gamma_{t\gamma}}{\Gamma_t}\times\frac{d\Gamma_{\bar{t}\gamma}}{\Gamma_t} \right)}_{\sigma_{\rm Decay}}\,,
\end{split}
\end{equation}
where the {\it Prod.} ({\it Decay}) contribution refers to the case where both photons are emitted in the production (decay) stage and the {\it Mixed} contribution contains all cases where we encounter photons simultaneously in the production and decay processes.

The dominant contribution of the $pp\to t\bar{t}\gamma(\gamma)$ process at LO originates from the QCD production of a top-quark pair at $\mathcal{O}(\alpha_s^2 \alpha^{4+n_{\gamma}})$, which we call \LOone, where $n_{\gamma}$ is the number of photons in the Born process. Additional power suppressed contributions are present at LO at $\mathcal{O}(\alpha_s^1 \alpha^{5+n_{\gamma}})$ (\LOtwo) and $\mathcal{O}(\alpha^{6+n_{\gamma}})$ (\LOthree) due to photon initiated subprocesses, the EW production of a top-quark pair and its interference to the QCD production. The sum of all three contributions is simply called \LOfull. The NLO corrections are dominated by the NLO QCD corrections to the LO QCD production (\LOone) at $\mathcal{O}(\alpha_s^3 \alpha^{4+n_{\gamma}})$, which is labelled \NLOone. The \NLOqcd calculation is then given by the sum of the two contributions, ${\rm NLO}_{\rm QCD}={\rm LO}_1+{\rm NLO}_1$,
which is then used in the following to quantify the size of the different resonant contributions according to Eq. \eqref{eq_res} in the $pp\to t\bar{t}\gamma\gamma$ process and to quantify the importance of the subleading contributions at LO and NLO.
\begin{figure}[t!]
  \begin{center}
  \includegraphics[trim= 20 580 20 20, width=0.5\textwidth]{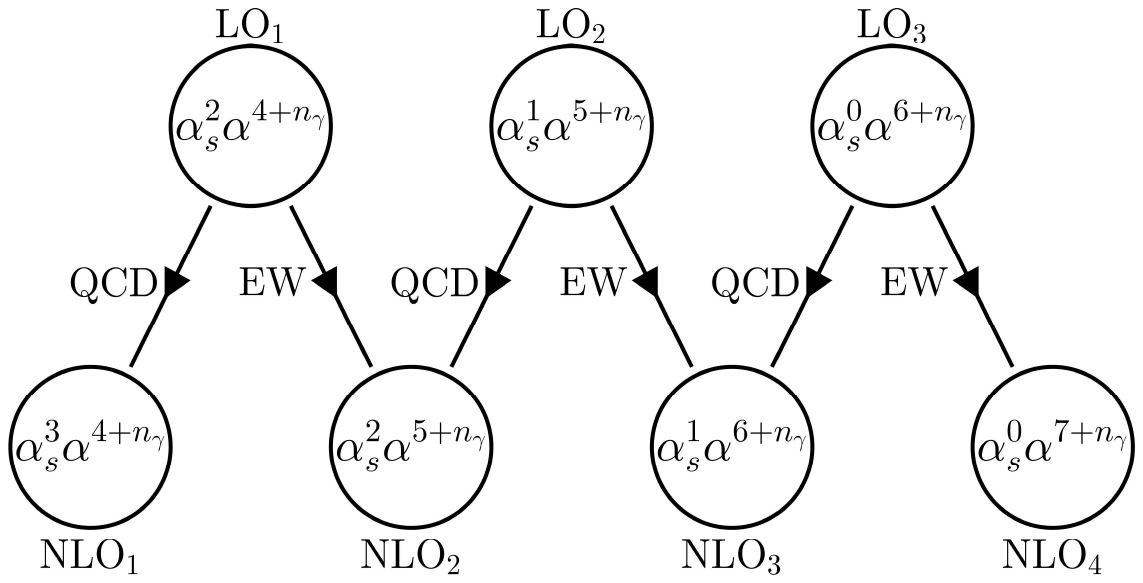}
\end{center}
  \caption{\label{fig:lo_nlo_cont} \it Interplay of LO and NLO contributions for the $pp\to t\bar{t}\gamma(\gamma)$ process with $n_\gamma=1(2)$. Figure was taken from \cite{Stremmer:2024ecl}.}
\end{figure}
As illustrated in Figure \ref{fig:lo_nlo_cont}, we have three additional subleading contributions at NLO due to QCD and EW corrections to different LO contributions. Similar to the LO contributions we label them as \NLOtwo, \NLOthree and \NLOfour. The complete \NLOfull calculation is then given by the sum of all LO and NLO contributions as
\begin{equation} \label{eq:nlo_tot}
{\rm NLO}={\rm LO}_1+{\rm LO}_2+{\rm LO}_3+{\rm NLO}_1+{\rm NLO}_2+{\rm NLO}_3+{\rm NLO}_4.
\end{equation}
In addition, we introduce an approximation, \NLOprd, in which we take into account exactly all LO contributions and \NLOone, but we neglect photon radiation and higher-order QCD and EW corrections in the decay processes in all subleading contributions. Thus, \NLOprd can be schematically written as
\begin{equation}
{\rm NLO}_{\rm prd}={\rm LO}_1+{\rm LO}_2+{\rm LO}_3+{\rm NLO}_1+{\rm NLO}_{2,{\rm prd}}+{\rm NLO}_{3,{\rm prd}}+{\rm NLO}_{4,{\rm prd}},
\end{equation}
where the subscript \textit{prd} indicates that photon radiation and higher-order corrections are taken into account in the production process only. This approximation does not only simplifies the calculation itself a lot, especially the real corrections, but also the matching to parton showers.

The general framework of our calculation is described in detail in Ref. \cite{Stremmer:2024ecl} which consists of the matrix element generator \Recola \cite{Actis:2016mpe,Actis:2012qn} with the program Collier \cite{Denner:2016kdg} for the calculation of scalar and tensor one-loop integrals, where we have further implemented the random polarisation method \cite{Draggiotis:1998gr,Draggiotis:2002hm,Bevilacqua:2013iha}. In addition, we have implemented a second method for the reduction of one-loop amplitudes using the OPP reduction method \cite{Ossola:2006us} with the program \textsc{CutTools} \cite{Ossola:2007ax} and the one-loop library \cite{vanHameren:2010cp}. The phase-space integration is performed with \textsc{Parni} \cite{vanHameren:2007pt} and \textsc{Kaleu} \cite{vanHameren:2010gg}. The calculation of the real corrections is carried out with the Nagy-Soper subtraction scheme \cite{Bevilacqua:2013iha} as implemented in \textsc{Helac-Dipoles} \cite{Czakon:2009ss}, which was recently extended for calculations with internal on-shell resonances for arbitrary QCD and QED-like singularities \cite{Stremmer:2024ecl}.

\section{Prompt photon distribution in \texorpdfstring{$pp\to t\bar{t}\gamma\gamma$}{ttaa}} \label{sec:phot_dist}
\begin{figure}[t!]
    \begin{center}
        \includegraphics[width=0.4\textwidth]{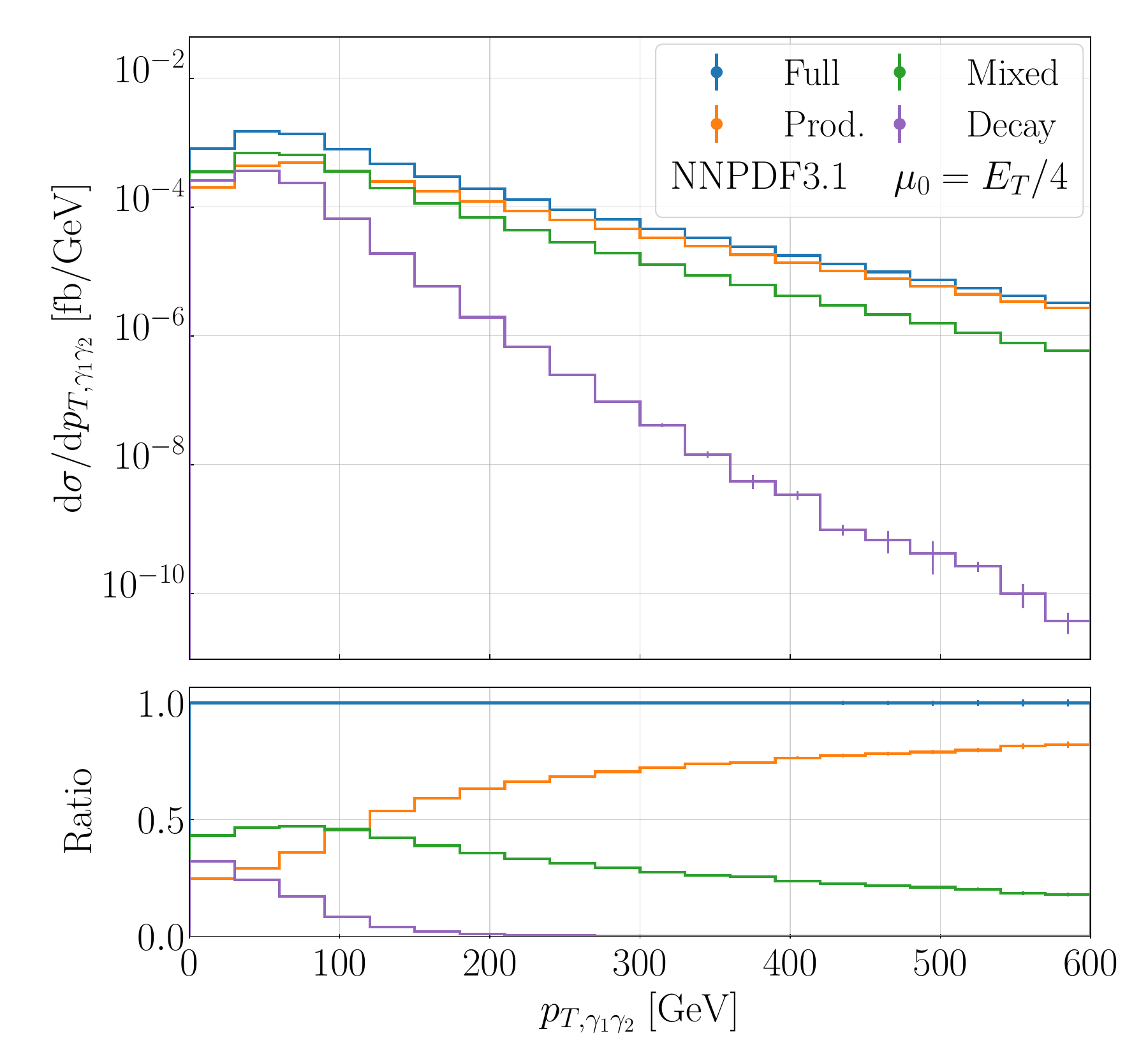}
        \includegraphics[width=0.4\textwidth]{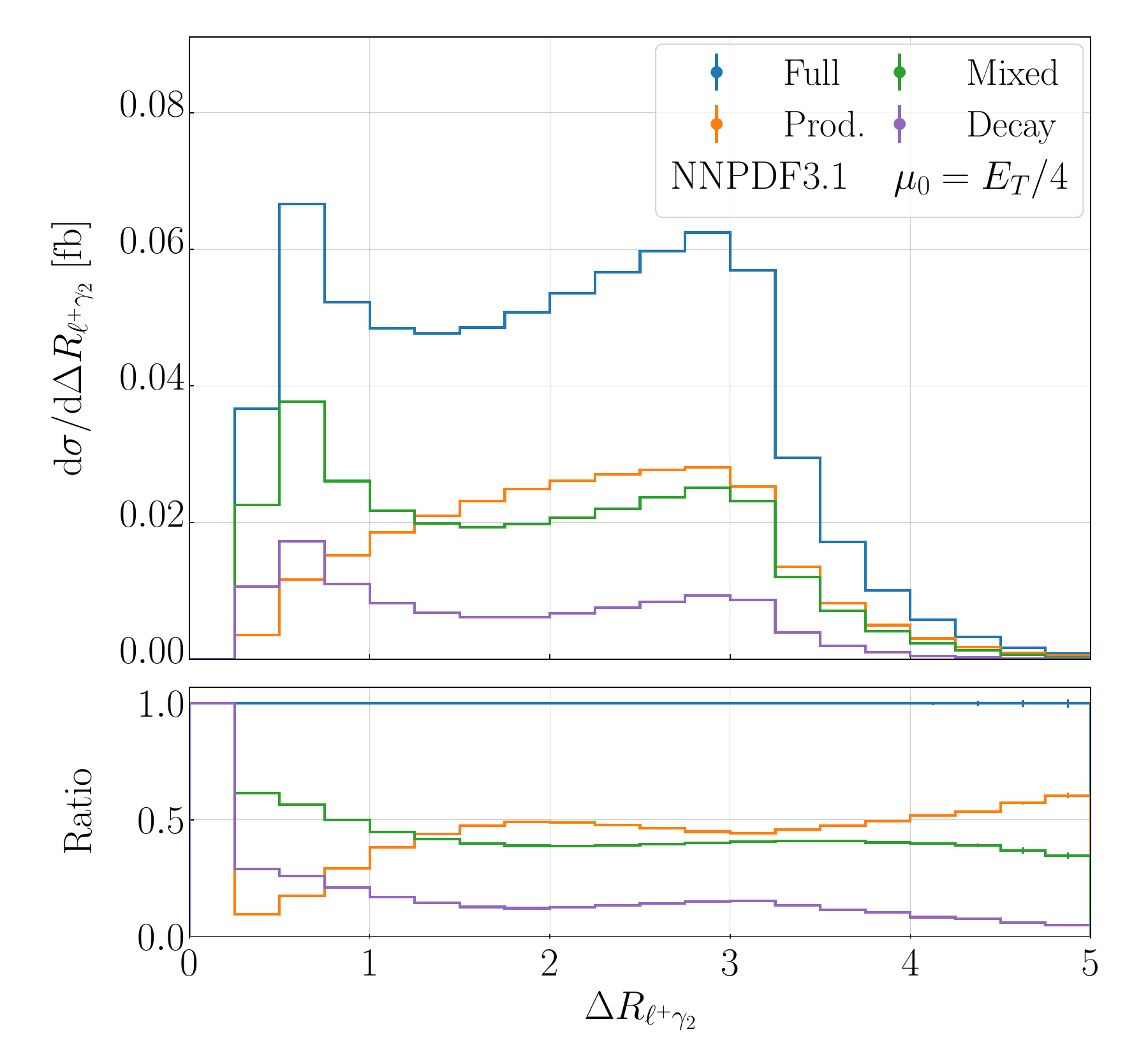}
    \end{center}
\vspace*{-6mm}
\caption{\label{fig:phot_dist} 
\it Differential cross-section distributions for the observables $p_{T,\gamma_1\gamma_2}$ and $\Delta R_{\ell^+\gamma_2}$ for the $pp\to t\bar{t}\gamma\gamma +X$ process in the di-lepton decay channel at \NLOqcd. Results are presented for the full process and the three resonant contributions {\it Prod.}, {\it Mixed} and {\it Decay}. The lower panels display the ratio to the full calculation. Figures were taken from \cite{Stremmer:2023kcd}.}
\end{figure}
The event selection and input parameters used for the following results are described in detail in Refs. \cite{Stremmer:2023kcd,Stremmer:2024ecl}. First we discuss the prompt photon distribution in the $pp\to t\bar{t}\gamma\gamma$ in di-lepton decay channel process at \NLOqcd and present in Figure \ref{fig:phot_dist} the observables $p_{T,\gamma_1\gamma_2}$ and $\Delta R_{\ell^+\gamma_2}$. At the integrated level, the {\it Prod.} contribution is only $40\%$ of the full calculation. Thus, the cross section is increased by a factor of $2.5$ when photon radiation is consistently included in the decays. The observable $p_{T,\gamma_1\gamma_2}$ displays the general behaviour of the three resonant contributions for dimensionful observables. In particular, at small energies the {\it Mixed} contribution is the dominant one with about $50\%$ of the full calculation while the {\it Decay} contribution can be as large as the {\it Prod.} contribution but rapidly decreases in size towards the tail. On the other hand, the {\it Prod.} contribution increases towards the tail and becomes the most dominant one with about $80\%$, while the {\it Mixed} contribution is still non-negligible with about $20\%$. The importance of photon radiation in the decays is even more pronounced for angular distributions such as $\Delta R_{\ell^+\gamma_2}$, where we find that the three resonant contributions have different peak structures. In particular, all three contributions are enhanced in the back-to-back region at $\Delta R_{\ell^+\gamma_2}\approx 3$, while the {\it Mixed} and {\it Decay} contributions have a peak at small angular separations, due to photon radiation in the decay processes, which is completely absent in the {\it Prod.} contribution.

\section{Complete NLO corrections} \label{sec:cnlo}

\begin{table*}[t!]
    \centering
    \renewcommand{\arraystretch}{1.1}
    \resizebox{0.65\columnwidth}{!}{
    \begin{tabular}{ll@{\hskip 10mm}l@{\hskip 10mm}l@{\hskip 10mm}}
        \hline
        \noalign{\smallskip}
         &&$\sigma_{i}$ [fb] & Ratio to ${\rm LO}_1$  \\
        \noalign{\smallskip}\midrule[0.5mm]\noalign{\smallskip}
        ${\rm LO}_1$&$\mathcal{O}(\alpha_s^2\alpha^5)$& $ 55.604(8)^{+31.4\%}_{-22.3\%} $ & $ 1.00 $ \\
        ${\rm LO}_2$&$\mathcal{O}(\alpha_s^1\alpha^6)$& $ 0.18775(5)^{+20.1\%}_{-15.4\%} $ & $ +0.34\% $ \\
        ${\rm LO}_3$&$\mathcal{O}(\alpha_s^0\alpha^7)$& $ 0.26970(4)^{+14.3\%}_{-16.9\%} $ & $ +0.49\% $ \\
        \noalign{\smallskip}\hline\noalign{\smallskip}
        ${\rm NLO}_1$&$\mathcal{O}(\alpha_s^3\alpha^5)$& $ +3.44(5) $ & $ +6.19\% $\\
        ${\rm NLO}_2$&$\mathcal{O}(\alpha_s^2\alpha^6)$& $ -0.1553(9) $ & $ -0.28\% $\\
        ${\rm NLO}_3$&$\mathcal{O}(\alpha_s^1\alpha^7)$& $ +0.2339(3) $ & $ +0.42\% $\\
        ${\rm NLO}_4$&$\mathcal{O}(\alpha_s^0\alpha^8)$& $ +0.001595(8) $ & $ +0.003\% $\\
        \noalign{\smallskip}\hline\noalign{\smallskip}
        ${\rm LO}$&& $ 56.061(8)^{+31.2\%}_{-22.1\%} $ & $ 1.0082 $ \\
        ${\rm NLO}_{\rm QCD}$&& $ 59.05(5)^{+1.6\%}_{-5.9\%} $ & $ 1.0620 $ \\
        ${\rm NLO}_{\rm prd}$&& $ 59.08(5)^{+1.5\%}_{-5.9\%} $ & $ 1.0626 $ \\
        ${\rm NLO}$&& $ 59.59(5)^{+1.6\%}_{-5.9\%} $ & $ 1.0717 $ \\
        \noalign{\smallskip}\hline\noalign{\smallskip}
    \end{tabular}
    }
    \vspace*{-2mm}
    \caption{\label{tab:tta} \it Integrated fiducial cross section for the $pp\to t\bar{t}\gamma +X$ process in the di-lepton top-quark decay channel. Results are presented at \LOfull, \NLOfull, \NLOqcd and \NLOprd with the corresponding scale uncertainties and for the individual LO and NLO contributions. Table was taken from \cite{Stremmer:2024ecl}.}
\end{table*}
In the following we concentrate on the $pp\to t\bar{t}\gamma$ process, but the conclusions are the same for $pp\to t\bar{t}\gamma\gamma$. The results of the \LOfull, \NLOqcd, \NLOfull and \NLOprd calculations at the integrated level  with the corresponding scale uncertainties as well as the individual (subleading) ${\rm LO}_i$ and ${\rm NLO}_i$ contributions are shown in Table \ref{tab:tta}. The \LOone and \NLOone contributions are the dominant ones at LO and NLO, respectively, while the subleading contributions are individually below $1\%$. Thus, the \NLOqcd calculation is completely sufficient to accurately describe this process at the integrated level and only differs by about $1\%$ with the complete \NLOfull calculation, while the corresponding scale uncertainties are about $6\%$.

\begin{figure}[t!]
    \begin{center}
        \includegraphics[width=0.4\textwidth]{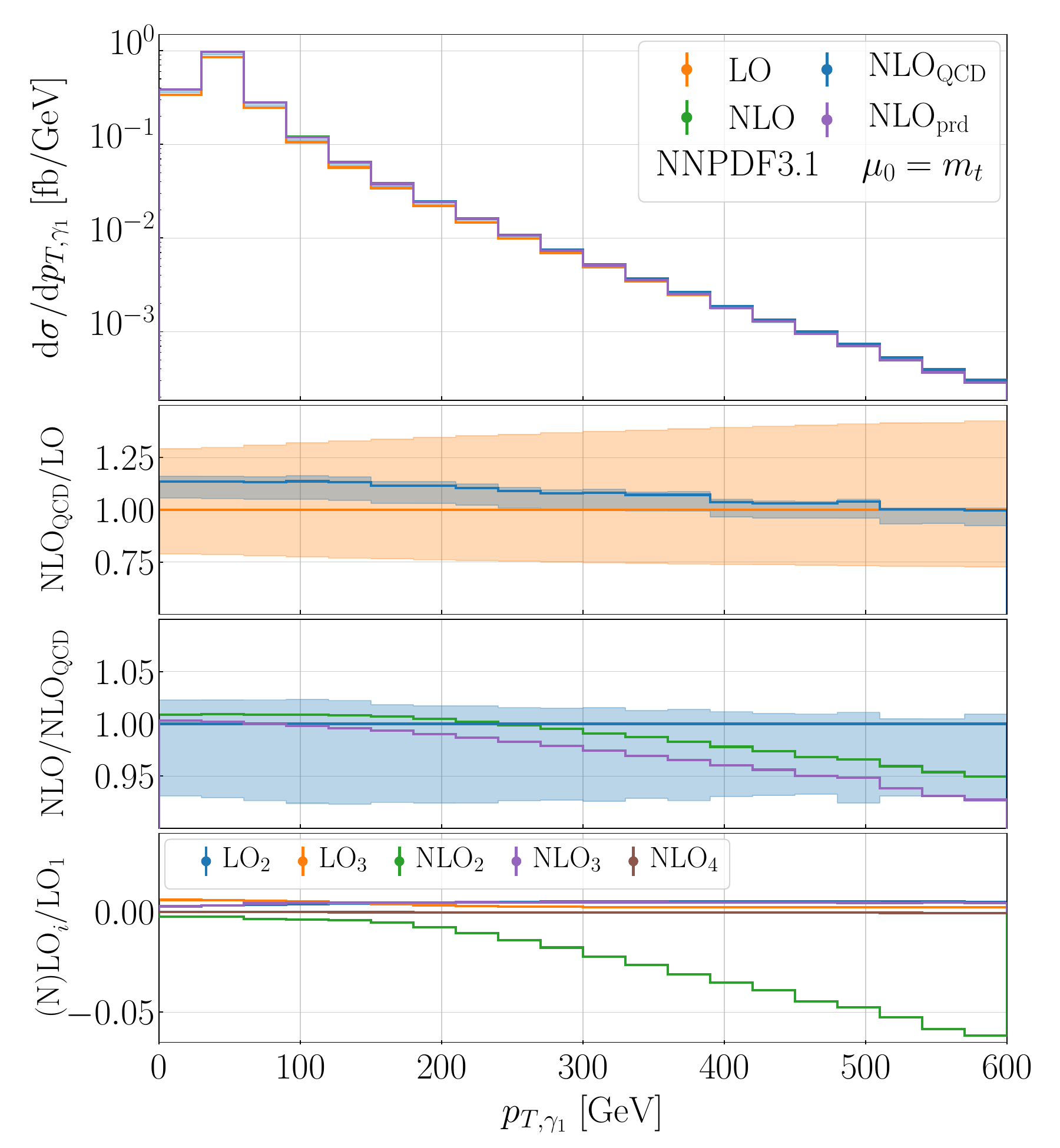}
        \includegraphics[width=0.4\textwidth]{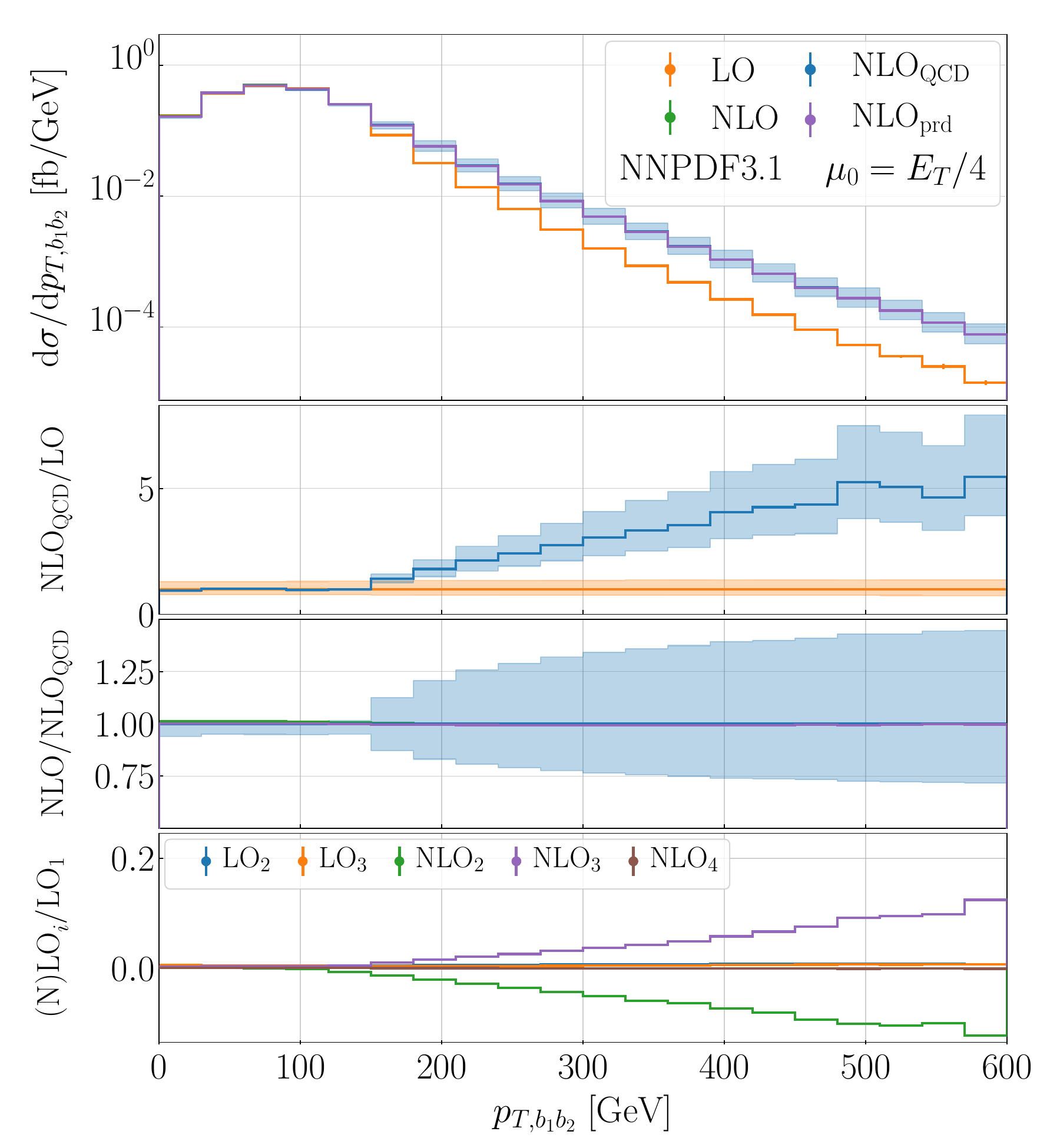}
    \end{center}
\vspace*{-6mm}
\caption{\label{fig:cnlo_tta} 
\it Differential cross-section distributions for the observables $p_{T,\gamma_1}$ and $p_{T,b_1b_2}$ for the $pp\to t\bar{t}\gamma +X$ process in the di-lepton decay channel. Results are shown at \LOfull, \NLOfull, \NLOqcd and \NLOprd. The second panels display the ratio of \NLOqcd with respect to \LOfull with the corresponding scale uncertainties. The ratio \NLOfull over \NLOqcd is shown in the third panels and the ratios of subleading LO and NLO contributions compared to \LOone are given in the lower panels. Figures were taken from \cite{Stremmer:2024ecl}.}
\end{figure}
In Figure \ref{fig:cnlo_tta} we present the differential results for the complete \NLOfull calculation in comparison to the \LOfull, \NLOqcd and \NLOprd calculations for the two observables $p_{T,\gamma_1}$ and $p_{T,b_1b_2}$. The lower panels display the ratios \NLOqcd over \LOfull, \NLOfull over \NLOqcd and the subleading (N)LO contributions over \LOone. We find that the largest subleading effects originate from the \NLOtwo contribution, due to EW Sudakov logarithms, which can be as large as $5\%-10\%$ compared to \LOone in the high-energy tails. This leads to a reduction of about $5\%$ of \NLOfull with respect to \NLOqcd for $p_{T,\gamma_1}$. On the other hand, both calculations are basically identical for $p_{T,b_1b_2}$ because of accidental cancellations between \NLOtwo and \NLOthree. The latter contribution is enhanced in the tail due to additional QCD radiation, where a similar pattern can already be found for the \NLOqcd calculation due to large real corrections in \NLOone. The differences between \NLOprd and \NLOfull are at most $2\%$ and thus significantly smaller than the corresponding scale uncertainties of about $8\%$ in this case.

\section{Conclusion}
In this report we have presented the prompt photon distribution in the $pp\to t\bar{t}\gamma\gamma$ process in the dilepton decay channel at NLO QCD, where we have demonstrated that the inclusion of photon radiation is crucial for a precise description, since the {\it Prod.} contribution is only $40\%$ of the full calculation at the integrated level. At the differential level, the situation is even more pronounced at small energies or for angular distributions such as $\Delta R_{\ell^+\gamma_2}$. On the other hand, the {\it Prod.} contribution dominates in high-energy tails, but the {\it Mixed} contribution remains non-negligible with about $20\%$.

In the second part of this contribution we discussed the calculation of subleading LO and NLO contributions for the $pp\to t\bar{t}\gamma(\gamma)$ process in the dilpeton decay channel. We have found that all subleading effects are negligible small at the integrated level and in sum amounts to $1\%$ with respect to the \NLOqcd calculation. The scale uncertainties are about $6\%$ and therefore significantly larger. On the other hand, the subleading contribution \NLOtwo is enhanced in high-energy tails of dimensionful observables due to EW Sudakov logarithms and can lead to a reduction of \NLOqcd by up to $5\%-10\%$. In addition, we have found accidental cancellations between \NLOtwo and \NLOthree, where the latter contribution is enhanced in the high-energy regime due to large real QCD corrections. Thus, the two contributions should only be considered together.  

\paragraph{Funding information}
This work was supported by the Deutsche Forschungsgemeinschaft (DFG) under grant 396021762 $-$ TRR 257: {\it P3H - Particle Physics Phenomenology after the Higgs Discovery}.

\end{document}